\documentstyle[12pt]{article}
\begin{document}
\def\V{{\cal V}}
\def\H{{\cal H}}
\def\C{{\cal C}}
\def\K{{\cal K}}

\pagestyle{empty}

\begin{flushright}
EFI 92-17 \\
\end{flushright}

\bigskip
\bigskip

\begin{center}

{\bf DISPERSION RELATIONS IN QUANTUM CHROMODYNAMICS}\footnote
{To appear in the  $\pi N$-NEWSLETTER No. 7} \\

\bigskip
\bigskip

{\it Reinhard Oehme } \\

\medskip
{\it Enrico Fermi Institute and Department of Physics \\
University of Chicago, Chicago, Illinois, 60637 } \\
\end{center}

\bigskip

\medskip

\bigskip

\centerline{ABSTRACT}

\begin{quotation}

  Dispersion relations for the scattering of hadrons are considered within the
framework of Quantum Chromodynamics.
It is argued that the original methods of proof remain applicable.
The setting and the spectral conditions are provided by an appropriate use
of the BRST\--cohomology.
Confinement arguments are used in order to exclude quarks and gluons from
the physical state-space.
Local, BRST\--invariant hadron fields are considered as leading terms in
operator product expansions for products of fundamental fields.
The hadronic amplitudes have neither ordinary nor anomalous
thresholds which are directly associated with the underlying
quark-gluon-structure.
Proofs involving the Edge of the Wedge Theorem and analytic completion
are discussed briefly.

\end{quotation}

\newpage

\baselineskip 20 pt
\pagestyle{plain}
\bigskip
\leftline{\bf 1. Introduction}
\smallskip

Dispersion relations for the scattering of hadrons have been
formulated [1-3] and proved in the Fifties [4-9]. They are by no means
simple generalizations of the familiar Kramers-Kronig relations
for the scattering of light [10,11]. The presence of finite masses
presents a formidable problem for obtaining the required
analytic continuations. Charges of various kinds give rise
to non-trivial crossing properties, which lead to analytic
connections of amplitudes for quite different reactions.

Even though they have been introduced a long time ago, dispersion
relations have continued to play an important r\^{o}le in the
analysis of hadron scattering. In a more general framework,
the analytic properties of Green's functions are the foundation
for many important results and theorems in field theory. However,
this analytic structure has not been discussed in detail within
the framework non-Abelian gauge theories like QCD and, in
particular, in the presence of confinement.

The original derivations of dispersion relations [4-9] are within
the framework of the general postulates of relativistic
quantum field theory of hadrons [12]. There is no need to
specify the theory in detail. The essential input is locality,
in the form of the existence of Heisenberg field operators,
which commute or anti-commute at space-like separations, and
which interpolate between asymptotic fields describing
non-interacting physical hadrons. In addition, spectral conditions
are very important for the proof. The difficulties with multi-particle
intermediate states, and with the analytic structure of
the corresponding multi-particle amplitudes, are the main
reason for the limitations of general proofs in some interesting
cases.

It is the purpose of this note to consider hadronic dispersion
relations within the framework of Quantum Chromodynamics (QCD).
As a constraint system, this $ SU(3) $ gauge field theory of
color is best quantized with the help of the Becchi-Rouet-Stora-Tyutin
(BRST) symmetry [13] in a state-space $\V$ of indefinite metric,
and in a covariant gauge like the Landau gauge, for example [14].
A priori, the space $\V$ contains quanta like ghosts and longitudinal
and time-like gluons, which are unphysical even without confinement. Using
the nilpotent BRST operator $Q$, we can define a subspace of states
which satisfy $Q\Psi = 0$. This is the kernel $kerQ = \{\Psi : Q\Psi = 0,
{}~\Psi \in \V \}$ of the operator $Q$. For ghost number zero, the
space $kerQ$ can provide the basis for a physical subspace,
provided we have {\it completeness} of the BRST operator [15]. This
notion implies that all states $\Psi \in kerQ$ {\it with zero norm}
are of the form $\Psi = Q\Phi$, $\Phi \in \V $. It is then easy
to show that $kerQ$ contains no states with negative norm. We can
define a cohomology space $\H = kerQ/imQ $ with zero
ghost number, containing only states with positive definite
norm. All zero norm states in $kerQ$ are contained in the subspace
$imQ = \{\Psi :  \Psi = Q\Phi, \Phi \in \V \}$. As is well known,
the cohomology space $\H$ provides a Lorentz-invariant definition
of a physical state-space.

Without completeness, states in $kerQ$ with zero norm and zero ghost number
could be made from ghosts and their conjugates (singlet pair
representations of the BRST algebra) [14,16]. There are arguments for
completeness [17,18], but we do not know of a general  proof for
four-dimensional gauge theories like QCD. Unless we have completeness, a
consistent formulation of the theory seems to be impossible.
In certain string theories, completeness has been proven explicitly
[15,19,20].

In the Hilbert space $\H$, the ghosts, as well as the
longitudinal and the time-like gluons, are eliminated in a
kinematical fashion, and
this is all that happens in weak coupling QCD perturbation theory.
But in the full theory, we expect that all quarks and gluons are
confined. With certain limitations concerning the number of
quark flavors, one can give arguments that,
for dynamical reasons, transverse gluon states cannot
be elements of the cohomology space $\H$ [21,22].
Some more preliminary arguments also exclude quark states [23].
Under these circumstances, $\H$ is a true
physical Hilbert space containing only hadronic states.
Here we adopt this algebraic view of confinement. It is quite
consistent with more intuitive pictures for the quark-gluon
structure of hadrons. In particular, the
existence of an approximately linear quark-antiquark potential
follows from the same arguments in a natural fashion [24,25].
Our arguments for confinement make use of the
renormalization group, and are valid for zero temperature.
At finite temperature, a new, dimensionful parameter comes in,
and there may be deconfinement transitions.

We assume here that exact QCD exists as a quantum field theory,
or that possible embeddings in more comprehensive schemes
are not of importance for confinement and for scattering
processes at energies well below the Planck mass. If local
field theory is considered as a low energy limit of string
theory, we may expect deviations from microscopic causality
at very small distances, and hence corresponding corrections to
dispersion relations.

Since the S-matrix, as an observable operator, is invariant under BRST-
transformations, it follows that the unitarity relations involve
only states from the subspace $\H$, at least as far as matrix
elements with respect to physical states are concerned [14,16]. With
the notion of confinement we have adopted, this implies that
only hadronic states play a direct r\^{o}le in the physical S-matrix.
Furthermore, intermediate state
decompositions of hadronic matrix elements of products of BRST-invariant
operators with zero ghost number require only a
complete set of hadronic states which
span the space $\H$. For the purpose of deriving hadronic
dispersion relations, this implies that the spectral conditions
remain the same as in the old hadronic field theory.
Of course, we assume here that there exist composite hadron
states in QCD.

As we have mentioned, the locality of the Heisenberg field
operators is the basis
for obtaining analytic properties of scattering amplitudes.
These operators interpolate between asymptotic fields,
which generate states of non-interacting particles [26].
Since we are interested in hadrons, we need to construct
local operators related to these particles in terms of quark
and gluon fields, which are the fundamental fields of QCD.
These hadronic, composite Heisenberg fields are BRST-invariant,
and they are asymptotically related to the corresponding
non-interacting hadron fields. They can be obtained, under
certain conditions, from the leading terms in the operator product
expansion [27] for the product of quark and antiquark operators
(mesons), or for three quark operators (baryons). The construction
is not unique, but there are equivalence classes  of interpolating
fields which give rise to the same S-matrix, as in the case of
fundamental fields.

Local field operators associated with the center-of-mass motion
of composite particles have been discussed extensively in the
literature [28-30]. It should not be surprising that such fields exist,
because locality does not imply a point-like structure for
the corresponding particles. In quantum electrodynamics, the
electron has charge and magnetic moment distributions, which
are described by the familiar form factors. Generally, in a relativistic
field theory, a given particle can be considered as a composite
of an appropriate set of the other particles.
The composite structure manifests itself in the form of
characteristic branch points for vertex functions and scattering
amplitudes. These structure singularities are caused by thresholds in
crossed channels of other amplitudes which are related by
unitarity to the amplitude under consideration [31].
For loosely bound systems, the structure branch points can
appear as anomalous thresholds [7,31-34] in the physical sheet of the
relevant variable.

In the case of
hadrons in QCD, a new element comes in. We do not have ordinary
composite systems where the constituents are observable
particles, but we have confinement. In principle, we may
consider a picture where nucleons and mesons are made up
of quarks with rather large constituent masses. But the
relevant quark masses appearing in QCD are the current masses,
which are very small in comparison with hadron masses, at
least as far as u- and d-quarks are concerned. The constituent
masses would have to be viewed as generated in connection
with the confinement process. In this process, gluons play
an important r\^{o}le. Formally, they come into consideration
via the anomaly in the trace of the energy-momentum tensor.
Estimates indicate, that the gluons actually give the dominant
contribution to the nucleon mass [35].
In view of the situation as described,
we cannot use weak-coupling perturbation theory in order to
argue for the existence of composite operators for the hadrons,
but we must rely on what is known in general about operator
products in local field theories [36].

As far as structure singularities and corresponding,
possible anomalous thresholds are concerned, the situation
in the case with confinement also differs from the
conventional picture of a composite system with observable
constituents. As explained above, hadronic amplitudes,
as expressed in terms of local hadronic fields, have
no thresholds associated with quarks or gluons.
Consequently there are no non-hadronic structure
singularities.

In deriving hadronic dispersion relations on the basis of
local, hadronic field theory, we have generally considered
the asymptotic condition as an additional postulate. We
do the same in QCD, where we use the condition essentially
only in the physical subspace. Since the theory involves gauge fields,
there is, a priori, no mass gap, and in its covariant form,
QCD operates in a state-space of indefinite metric. The
Haag-Ruelle arguments [37] for obtaining the asymptotic
condition from the other postulates of the theory
are not applicable under these circumstances.

In the following Section we give a brief account of the
hadronic subspace and of confinement. In Section 3, we
discuss some relevant aspects
of composite operators and of operator product expansions.
Section 4 is devoted to a very brief survey of the analytic methods
used in proving dispersion relations.

This article can give only a brief sketch of the many problems
involved in deriving analytic properties of hadronic amplitudes
in QCD. We hope to present a more comprehensive report elsewhere.
\newpage
\bigskip
\bigskip

\leftline{\bf 2. Hadronic Subspace}
\smallskip

In this Section we give a brief introduction to the construction
of a physical subspace with positive definite metric, which, in
view of confinement, can serve as a space spanned by hadronic
states exclusively. Under these circumstances,
the spectral conditions used in the
derivation of dispersion relations for hadrons remain the same
as in the generic, hadronic field theory used in the past.

As explained in the Introduction, we consider
QCD in a covariant gauge, and quantize in a space $\V$ of
indefinite metric in accordance with BRST-symmetry. The
self-adjoint BRST operator $Q$, and the corresponding ghost
number operator $Q_c$, form the algebra [13,14]
$$
Q^2 = 0, ~~ i[Q_c,Q] = Q~,
\eqno (2.1)
$$
which can be used to generate a decomposition of $\V$ in
the form
$$
\V = kerQ \oplus \V_u ~, ~~ kerQ = \V_p \oplus imQ ~.
\eqno (2.2)
$$
Here we have introduced the subspaces
$$
kerQ = \{\Psi : ~Q\Psi = 0, ~ \Psi \in \V \}~,
\eqno (2.3)
$$
and
$$
imQ = \{\Psi : ~ \Psi = Q\Phi, ~ \Phi \in \V \}~.
\eqno (2.4)
$$
We notice that $imQ \perp kerQ $ with respect to the indefinite
inner product $(\Psi,\Phi)$ defined in $\V$. A priori, the
subspace $\V_p$ is a candidate for a physical statespace, but
it is not invariant under Lorentz transformations, nor under
equivalence transformations, which leave the physics unchanged.
As is well known, one therefore uses the cohomology space
$\H = kerQ/imQ$, which is a space of equivalence classes.
It is isomorphic with $\V_p$. A state $\Psi \in \H$ may be
written symbolically as $\Psi = \Psi_p + imQ,~ \Psi_p \in \V_p$~.
We have ignored here the grading due to the ghost number
operator, since we are interested in the sector
$N_c = 0$ as far as $\H$ is concerned.

In order to assure a physical subspace $\H \simeq \V_p$
with positive definite metric, we must
assume {\it completeness} of the BRST operator $Q$ [15]. This notion
implies that all states with zero norm in $kerQ$ are contained in $imQ$.
Given completeness, it is easy to see that there cannot be
any negative norm states in $kerQ$. It is not enough
to have zero ghost number, because the singlet pair representations
of the BRST algebra (2.1), which include states of ghosts and
anti-ghosts, must also be eliminated. There are arguments for
the absence of singlet pairs in the dense subspace generated
by Heisenberg operators, but in view of the indefinite metric,
the extension to the full space $\V$ is delicate [17,18]. In certain
string theories, completeness has been proven explicitly.
In any case, without completeness of the BRST operator,
a consistent formulation of QCD would seem to be impossible.
{}From a mathematical point of view, the actual sign of the definite
norm in the cohomology space is a convention.

Given completeness, we use a simple matrix notation
for the zero ghost number sector of $\V$, with components
referring to the subspaces $\V_p$, $imQ$ and $\V_u$ respectively.
We write
$$
\Psi = \left ( \begin{array}{c} \psi_1 \\ \psi_2 \\ \psi_3
\end{array}\right), ~~  \C = \left( \begin{array}{clcr}
1 & 0 & 0  \\
0 & 0 & 1  \\
0 & 1 & 0
\end{array} \right) ,
\eqno (2.5)
$$
where the self-adjoint involution $\C$ may be viewed as a
metric matrix. In terms of components, the inner product in $\V$
is then given by
$$
(\Psi ,\Phi ) = (\Psi , \C\Phi )_{\C} =
\psi_1^* \phi_1 + \psi_2^* \phi_3 + \psi_3^* \phi_2~,
\eqno (2.6)
$$
where the subscript $\C$ denotes an ordinary inner product.
We see that for states $\Psi , \Phi \in kerQ $, which are
representatives of physical states, only the first term on
the right-hand side of Eq.(2.6) remains. Since $\V_p$ is a
non-degenerate subspace, we can define a projection operator
$P(\V_p)$ with $P^\dagger = P^2 = P$. For the inner product
of two states $\Psi , \Phi \in kerQ$, and with a complete set
of states $\{\Psi_n\}$ in $\V$, we obtain then the decomposition
$$
(\Psi , \Phi ) = \sum_n (\Psi ,\Psi_n )(\Psi_n ,\Phi)  \\ =
(\Psi , P(\V_p) \Phi ) = \sum_n (\Psi , \Psi_{pn})
(\Psi_{pn} , \Phi )~.
\eqno (2.7)
$$
We see that only a complete set of states $\{\Psi_{pn}\}$ in
the Hilbert space $\V_p \simeq \H$ appears in the sum.
It may be replaced by the equivalent set $\{\Psi_{\H n}\}$,
where we can write symbolically $\Psi_{\H n} = \Psi_{pn} +
imQ$. Although the projection operator $P(\V_p)$ is not Lorentz
invariant by itself, the use in Eq.(2.7) is invariant.

In our matrix representation, a BRST-invariant operator A,
which commutes with $Q$, and leaves $kerQ$ as well as $imQ$
invariant, is of the form
$$
A = \left ( \begin{array}{clcr}
A_{11} & 0 & A_{13}  \\
A_{21} & A_{22} & A_{23}  \\
0      & 0      & A_{33}
\end{array} \right) .
\eqno (2.8)
$$
Given a state  $\Psi \in kerQ$, it follows that also $A\Psi \in kerQ$.
With Eq. (2.7), states $\Psi , \Phi \in kerQ$, and BRST invariant
operators $A$ and $B$, we have therefore the decomposition
$$
(\Psi , AB \Phi ) = \sum_n (\Psi , A\Psi_{pn} )(\Psi_{pn} ,B\Phi )~,
\eqno (2.9)
$$
which involves only physical states [14,21,38,39].

Proper unphysical states  are characterized by $Q\Psi \neq 0$, or
$\psi_3 \neq 0$ in our matrix representation (2.5). They
may well have components in $\V_p$, but one can always find an
equivalence transformation which removes this component [38,39].
In general, the norm of these states is indefinite. It can be shown, that
some unphysical states with positive norm are needed for the consistency
of the theory [21].

In QCD perturbation theory, the space $\H$ consists of states
corresponding to quarks and transverse gluons. But in the general
theory, if we have confinement, only colorless states like
hadrons should be included. We understand here confinement in
this algebraic fashion. If the number of flavors $N_F$ in QCD
is limited ($N_F \leq 9 $), arguments can be given that gluons
cannot be in the physical subspace [21,22]. These arguments are based
upon the existence of superconvergence relations for the
structure function of the gluon propagator, which provide a
connection between short and long distance properties.
Within the same framework, we can also obtain an approximately
linear quark-antiquark potential, because a dipole representation
can be written for the propagator, which has a weight function of
the appropriate shape [24]. While a linear potential may be a
phenomenological reality for heavy quarks, we require an algebraic
argument also for quark confinement. A sufficient condition
for general color confinement in terms of the BRST-cohomology
has been given by Kugo and Ojima [14], and
discussed further by Nishijima [40]. So far, only approximate methods
have been used in order to
argue that this condition is also necessary [23]. But if we accept
the necessity, the confinement of transverse gluons also
implies the confinement of quarks.

For the purpose of deriving dispersion relations in QCD, we take it for
granted that we have confinement in the sense that the
physical state-space $\H$ contains only hadronic states. Under these
circumstatences, quarks and gluons do not appear as BRST singlets, but form
quartet representations of the algebra, together with other
unphysical states.
As we have seen in Eq.(2.9), only hadronic states appear
then in intermediate state decompositions involving hadronic
operators, and we have the same spectral conditions as used in the old
derivations of dispersion relations within the framework of hadronic
field theory. Under these circumstances,
in the direct or crossed channels of amplitudes, there appear
no thresholds which are associated with the quark-gluon structure.
Also anomalous thresholds related to this substructure do not
exist, neither in the physical nor in the unphysical sheets of
the relevant variable, since they are generated by crossed
channel thresholds of amplitudes related by unitarity
to the one under consideration.\footnote{A detailed discussion
of the mathematical and physical aspects
of structure singularities may be found in Ref. 31.}
We have a situation, which is quite different
from the usual bound state system, where the constituents are
physical particles which can contribute to intermediate state
decompositions as in Eq.(2.7).

Although the quark-gluon structure of hadrons does not give
rise to special singularities of hadronic amplitudes,
it is expected to be of importance in the determination of the weight
functions (discontinuities) in dispersion representations.
To the extent that perturbative QCD is related to the
weak coupling limit ($g^2 \rightarrow +0$ , $g$ =
gauge coupling) of the full theory, we may expect to see
evidence of the composite structure in regions of
momentum space where the effective gauge coupling is small
as a consequence of asymptotic freedom. In these regions,
pertubation theory may be a reasonable tool for the
approximate determination of weight functions. Since there is
no confinement in perturbation theory, which is the extreme
asymptotic limit $g^2 \rightarrow +0$ ,
the absence of quarks and gluons from the physical space
$\H$ of the full theory should be related to the required
hadronization.

Our algebraic description of confinement should be in accordance
with other, perhaps more intuitive approaches to the problem.
{}From the point of view of covariant field theory, the mathematical
question is always whether or not a given excitation is in the
physical cohomology space $\H$. Only states in $\H$ are
observable and contribute directly to the singularity structure
of hadronic amplitudes.

\newpage

\bigskip

\bigskip
\leftline{\bf 3. Local Hadronic Operators}
\smallskip

The problem of dealing with composite particles in quantum
field theory was considered in the late Fifties [28,29]. The methods
can be generalized to gauge field theories with a state space
of indefinite metric. It is possible to define {\it local}
field operators for stable composite systems, which interpolate
between the asymptotic fields generating these bound states as incoming
or outgoing particles in a scattering process.

Let $\psi (x)$ describe fundamental fields of the theory.
We suppose that there exists a stable, composite system,
which has a rest-mass $M$, and quantum numbers in accordance
with those of a product like $\psi \overline{\psi}$. Then we may
associate a local operator field $B(x)$ with the composite system.
The field $B(x)$ could be defined as the limit
$$
B(x) = \lim_{\xi \rightarrow 0} \frac {\psi (x + \xi)
\overline{\psi} (x - \xi)}{F(\xi)},
{}~~ -{\xi}^2 < 0.
\eqno(3.1)
$$
The space-like approach is convenient, but not essential.
The invariant function $F(\xi)$ is only of importance as far as it's
behavior for $\xi \rightarrow 0$  is
concerned. Generally, the function is singular in this limit,
in order to compensate for the expected singularity of the
operator product. In fact, it must be as singular as the most
singular matrix elements of this product. Here we make the
assumption that such maximal matrix elements exist. Otherwise, we
would have a situation where, for every matrix element with
a given singularity, there exists a more singular one.
Given the existence of maximal matrix elements, we generally
have an equivalence class $\K_{max}$ of functions $F$ with
the required maximum singularity.
Possible oscillations in the limit (3.1) can be
handled by an appropriate choice of the sequence of points in
the approach to $\xi = 0$.

The limit (3.1) corresponds to the leading term in an
operator product expansion [27]. Such expansions are known to exist
in many lower-dimensional field theory models. They are
expected to be a general property of local field theories.
In four dimensions, the existence of operator product expansions,
and of local, composite operators like $B(x)$ in particular,
can be proven using perturbation theory methods of renormalizable
field theories [41]. But in the corresponding exact theories, we
still have to make the technical assumption concerning the
maximal singularity mentioned above [36]. As explained in the
introduction, we should not rely upon QCD perturbation theory
for the purpose of deriving hadronic dispersion relations.

Given the existence of local, BRST-invariant operators in QCD which are
associated with hadrons, we can write representations
for amplitudes as Fourier transforms of time ordered or
retarded products of these operators. The Fourier representations
are then the starting points for obtaining analytic properties.
In order to give some more details, we consider briefly an
amplitude for the elastic scattering of hadrons in QCD. For
simplicity, we ignore spin and other quantum numbers, concentrating
on the general structure of the S-matrix elements.
Consequently, the following formulae are rather symbolic.
Let us define time-ordered products of basic fields in the form
$$
B(x,\xi) = T \psi (x + \xi) \overline {\psi} (x - \xi) ,
\eqno (3.2)
$$
or
$$
B(x; \xi_1, \xi_2, \xi_3 ) = T \psi (x + \xi_1)
\psi (x + \xi_2) \psi (x + \xi_3) ,
\eqno (3.3)
$$
with $\xi^2 < 0$ and the distances $\xi_i - \xi_j$ kept space-like.
We assume that these operators have non-trivial hadronic
quantum numbers, so that their vacuum expectation value vanishes.

Considering  $B(x, \xi)$, we suppose that there exists a
hadron (meson) with mass $M$ so that
$\langle 0 | B(x, \xi) | k \rangle \neq 0 $
for $-k^2 = M^2$, where $ | k \rangle $ is the single hadron state.
The free retarded  and advanced propagator functions
$\Delta_{R,A} (x - x', M)$ can be used to define asymptotic
fields $B_{in}(x, \xi)$ and $B_{out}(x, \xi)$ with the help
of the Yang-Feldman representation. With
$$
\langle 0 | B(x, \xi) | k \rangle = \langle 0 | B_{in}(x, \xi) |k\rangle =
e^{ik\cdot x} F_k (\xi) ,
\eqno (3.4)
$$
we introduce a function  $F_k(\xi) = \langle 0 |B(0, \xi)|k\rangle$.
Denoting the Fourier transform of $B_{in}(x, \xi)$ by
$B_{in}(k, \xi)$, we can show that there are creation and
destruction operators like
$$
\frac{B^*_{in}(k,\xi)}{F_k(\xi)} = b^*_{in}(k) ,
\eqno (3.5)
$$
which are independent of $\xi$ and satify the usual commutation
relations. In this derivation, the completeness of states
in $\V$, which are generated by all asymptotic fields, including
composite fields, has been assumed [28,29].

In principle, we may consider asymptotic fields for unphysical
excitations in the state-space $\V$ of indefinite metric.
The states generated by these fields are not elements of
the physical space $\H$, the cohomology space of the BRST
oprator. We associate asymptotic fields with the poles
of time ordered Green's functions corresponding to non-negative
eigenvalues of $-P^2$, where $P$ is the energy-momentum tensor [14].
We do not exclude here the possibility of multipole fields.

With the asymptotic fields (3.5), and the weak asymptotic
condition
$$
\lim_{\xi \rightarrow 0} (\Psi, B^f (x^0 , \xi )\Phi ) =
(\Psi, B^f_{in}(\xi ) \Phi )
\eqno (3.6)
$$
for all $\Psi, \Phi \in \V $, where
$$
B^f(x^0, \xi ) = -i\int d^3 x B(x, \xi )\stackrel{\leftrightarrow}{\partial^0}
f^*(x)~,
\eqno (3.7)
$$
for any normalizable $f(x)$ satisfying $K_x f = (\Box - M^2)f(x) = 0$,
we can use the reduction formulae of Lehmann, Symanzik and
Zimmermann [26] in order to obtain representations for hadronic amplitudes.
For example, let us consider the scattering of mesons with mass
$M$. With the product of basic fields as defined in Eq.(3.2),
we obtain a formula like
%\newpage
$$
 \langle k',p' |S| k,p \rangle =
\frac{1}{F_{k'} (\xi ')
\overline{F}_k (\xi )}  \frac{1}{(2\pi )^3}  \int \int d^4 x' d^4 x
\exp[-ik'x'+ikx]
$$

$$
  K_{x'} K_x  \langle p' | T B(x',\xi ')
\overline{B}(x, \xi ) | p \rangle ,
\eqno (3.8)
$$
where $-k^2$ =  $-{k'}^2$ = $M^2$, and $ |p \rangle , ~ |p' \rangle $
are single hadron {\it in}-states.
The right-hand side of Eq. (3.8) is independent of the relative
coordinates $\xi $ and $\xi '$.

So far, we have not taken the limit $\xi ,\xi' \rightarrow 0$.
This limit is necessary in order to have the microscopic
causality required for dispersion relations. Furthermore,
the operator $B(x,\xi)$ is not BRST-invariant for $\xi \neq 0$.
Only a local limit like $B(x)$ in Eq.(3.1) is invariant. As suggested
by representations like Eq.(3.8), and the properties of the
operators $B(x, \xi )$, we restate the assumption made in
connection with Eq.(3.1) and suppose that the limit
$$
B(x) = \lim_{\xi \rightarrow 0} \frac {B(x,\xi )}{F_k (\xi )}
\eqno (3.9)
$$
exists. It then defines a local hadronic Heisenberg operator, and
it implies that the functions $F_k (\xi )$ are elements of the
equivalence class $\K_{max}$ mentioned above.
If we now interchange the local limit and the space-time integrations
in Eq. (3.8), we obtain a representation of the S-matrix element
in terms of local hadron fields :
%\newpage
$$
\langle k',p' |S | k,p \rangle = \frac {1}{(2\pi )^3} \int\int
d^4 x' d^4 x \exp [-ik'x'+ikx]
$$

$$
K_{x'} K_x \langle p' | T B(x') \overline {B}(x) | p \rangle .
\eqno (3.10)
$$
We can make further reductions in Eq.(3.10) in order to get the
formulae needed  for the derivation of non-forward dispersion
relations, and of forward relations for amplitudes with
unphysical continuum contributions.

Instead of taking the limit  $\xi \rightarrow 0$ in Eq.(3.8),
we can use directly the local operator (3.9) and its
asymptotic limit in order to construct the scattering
amplitude (3.10) with the help of the reduction formula
involving the local composite field $B(x)$. Under these circumstances, the
reduction method is used only within the physical subspace,
where there should be no problems resulting from the infra-red
singularities of the theory. But even though the path via
Eq.(3.8) appears to involve more assumptions, we think that
it may be of interest for the understanding of the hadronic,
local limit.

The reduction described above for the product (3.2) can be
generalized to operator products like (3.3), as well as
to other products of fundamental fields which can form color singlets. In this
connection, it is important to note that the Heisenberg
fields interpolating between given asymptotic, hadronic fields
are not unique. There are equivalence classes of fields giving
rise to the same S-matrix.

For field theories with a state space of positive definite metric,
it can be shown that locality is a transitive property:
two fields, which commute with a given local field, are local
themselves and with respect to each other. We have equivalence
classes of local fields (Borchers classes) [42]. The proof involves
the equivalence of weak local commutativity
and CPT-invariance [43], as well as the Edge of the Wedge Theorem [7].
It is then possible to show that different fields in a given
class, which have the same asymptotic fields, define the same
S-matrix.

Given special rules for the transformation of ghost fields
under CPT, we can define an anti-unitary CPT-operator
in the state-space $\V$ of QCD [14]. Together with the other
postulates of indefinite metric field theory, this then leads
to the existence of equivalence classes of local Heisenberg
fields in QCD. In particular, the hadron fields $B(x)$, defined
by different versions of the local limit, are in the
same equivalence class, as are corresponding products
involving basic fields as factors.
This is a consequence of the locality of the basic fields
in QCD. As long as the different composite fields $B(x)$ have
the same quantum numbers and the same {\it in}-fields, they
give rise to the same physical S-matrix in the subspace
$\H$ of hadron states.

\newpage

\bigskip

\bigskip
\leftline{\bf 4. Methods of Proof}
\smallskip

In previous Sections, we have explained that one can define
local Heisenberg Operators for hadrons in QCD. We have seen
that BRST methods allow for the definition of a physical
subspace $\H$ of the general state-space $\V$ of QCD.
Given confinement, the Hilbert space $\H$ contains only
hadronic states. With these features of QCD, we can proceed
to derive dispersion relations using the methods developed
within the general framework of local, hadronic field
theory. In the following, we recall briefly some of
the essential mathematical steps in the proof of dispersion
relations for forward scattering, and for finite values of
the momentum transfer.

Dispersion relations for the forward scattering amplitudes
of reactions like pion-pion scattering and pion-nucleon
scattering can be derived rather simply by using the
{\it gap method}.\footnote{The gap method was introduced
in Ref. 4 (see the appendix, in particular).} As a
simple model, let us consider the scattering of massive
scalar particles. In terms of local Heisenberg fields, the
forward amplitude has the representation

$$
F ( \omega ) = \int d^4 x e^{i \omega x^0 - i \sqrt{\omega^2 - \mu ^2}
\hat{e} \cdot \vec{x}} \theta (x^0) \chi (x^0 , | \vec{x} | ) ~~,
\eqno(4.1)
$$
where
$$
\chi (x^0 , | \vec{x} | ) =
\frac{i}{(2 \pi )^3} \langle p | \left [
j ( \frac{x}{2}), j ( - \frac{x}{2}) \right ] | p \rangle ~~,
\eqno(4.2)
$$
with $ j \equiv ( \Box - \mu^2 ) \phi$.

If we write, with $ r = |\vec{x}| $,
$$
F ( \omega ) = \int_0^\infty dr F (\omega , r ) ~~,
\eqno (4.3)
$$
we find that, for the relevant values of $r$, the function
$F (\omega , r )$ is analytic in the upper half of the complex
$\omega$-plane, because the integrand in Eq. (4.1) has
support only in the future cone.
As a Fourier transform of a tempered distribution,
it is bounded by a polynomial. We ignore a possible polynomial,
which can be taken care of by subtractions, and write a Hilbert
representation for $F(\omega , r)$. This representation involves an integral
with the weight function $Im F(\omega + i0 , r )$ along the
real $\omega$ - axsis. As a consequence of the spectral conditions,
the weight function vanishes in the gap $- \mu < \omega < + \mu $.
But for $ |\omega | \geq \mu $, we can perform the $r$ - integration
on both sides of the Hilbert transform. Using the crossing
symmetry for the neutral, scalar model, we obtain the dispersion
relation
$$
F ( \omega ) =
\frac{2 \omega}{\pi}\int_{\mu^2}^\infty d \omega '
\frac{Im F ( \omega '  + i 0 )}{{\omega '}^2 - \omega^2}~~.
\eqno(4.4)
$$

Although the arguments sketched above ignore many fine-points,
they show directly how locality (microscopic causality) and
simple spectral conditions translate into the analytic
properties required for the validity of dispersion relations.
The generalization of the gap method to cases with single
particle states, like the nucleon in pion-nucleon amplitudes,
is straightforward. We simply remove the one-particle
contribution by the appropriate factor, and later regain it
as a pole term in the once-subtracted  dispersion relation.
The real coefficient of the single-nucleon term can be
identified with the pion-nucleon vertex function on the
mass shell [5]. A proof involves  applying the gap method also to
this vertex function in a nucleon channel. For reactions involving
charged particles,
like $\pi^{\pm} p$ - scattering, we have non-trivial crossing
relations in the sense that the physical amplitudes for $\pi^+ p$ - and
$\pi^- p$ - scattering are different boundary values of the same
analytic function, which is regular in the cut $\omega$-plane,
except for the single nucleon pole.

For reactions like  $\pi\pi$- and $\pi N$- scattering, we can
prove near-forward dispersion relations with the help of the
gap method. These relations involve the derivatives of amplitudes
with respect to the momentum transfer $t$, evaluated at $t = 0$ [2].
But for amplitudes with fixed, finite momentum transfer, more
sophisticated methods must be used. In these cases, we have continuous
unphysical regions. The same is true for forward amplitudes
for reactions like nucleon-nucleon scattering, where the crossed
channel involves nucleon-antinucleon scattering, and has
continuous contributions from states with two or more mesons.

The natural mathematical framework for the derivation of
these dispersion relations is the theory of functions of
several complex variables. Two aspects of this theory are
of fundamental importance for our purpose: 1. The
{\it Edge of the Wedge Theorem}, and 2. the existence
of {\it Envelopes of Holomorphy}.

In order to describe the Edge of the Wedge Theorem [7], we use an example
involving one complex four-vector. Suppose amplitudes,
like those for $\pi^+ N$- and $\pi^- N$- scattering, are given
as Fourier transforms of tempered distributions with support
in the future or the past light-cone respectively:
$$
F_{\pm} (K) = \pm\frac{i}{(2\pi)^3} \int d^4 x~ e^{-i K \cdot x}
\theta (\pm x^0)
{}~\langle p' | \left [ j^{\dagger}(\frac{x}{2}), j(-\frac{x}{2} \right ]
| p \rangle~~.
\eqno (4.5)
$$
Here $2K = k + k'$, and $ k + p = k' + p' $.
As a consequence of locality, the functions $F_{\pm}(K)$ are
analytic in the tubes -$(ImK)^2 > 0$, $Im K^0 > 0$ or
$Im K^0 < 0$ respectively. For real values of the four-vector
$K$, outside of the physical regions for both reactions,
there is a domain $R$ where the two functions coincide. This is
a consequence of the spectral conditions, as may be seen by
making a decomposition of the absorptive parts with respect
to a complete set of hadron states. Given the situation as described,
the Edge of the Wedge Theorem implies that there exists a complex
neighborhood $N(R)$
of the real domain $R$, where both functions are analytic and
coincide. Hence there exists a unique analytic function $F(K)$,
which is regular, at least, in the union of $N(R)$ and the
region  -$(ImK)^2 > 0$.  It is important to note that $N(R)$ contains
all points
with  sufficiently small, {\it space-like} imaginary part,
which are not points of the original tubes.
The physical amplitudes $F_{\pm}(K)$ are boundary values of the
general analytic function $F(K)$ in the appropriate real regions.
In many cases, the domain of analyticity obtained from
the Edge of the Wedge Theorem is not yet large
enough for dispersion relations, but it gives an analytic
connection between the two physical amplitudes, and hence
a meaning to the crossing relations.

The case described above is only a simple example of the
Edge of the Wedge theorem. It has been generalized in many ways.\footnote
{The Edge of the Wedge Theorem has many applications beyond the
problem of dispersion relations. In the literature, one can find
elaborate explanations concerning the origin of the name. In fact,
while working on the problem in Princeton in 1956-57, we (BOT)
called it {\it Keilkanten Theorem}, which was simply translated
for the publication [7].}
For the proof of dispersion relations at fixed momentum transfer t, we
have used it for functions of two complex four-vectors,
together with an original domain of analyticity of the
form $W\otimes W$, where $W$ is the tube -$(ImK)^2 > 0$
used above [7].

The other essential tool for the derivation of non-forward
dispersion relations is analytic completion. For functions with
two or more complex variables, we have the remarkable situation
that, for many domains $D$, {\it all} functions, which are
holomorphic in $D$, can be continued into a larger domain $E(D)$,
the {\it envelope of holomorphy } of $D$. This envelope is a
purely geometrical notion. The basic, generic tool for the
construction of envelopes of holomorphy is the Continuity
Theorem, which has been used in Ref.44 to give a complete
construction of $E(W\cup N(R))$, where $W\cup N(R)$ is the
domain of analyticity described in the example for the Edge of
the Wedge theorem given above. On the other hand, in Ref.7,
we have used a subdomain, which is a generalized semitube, and
for which the envelope is well known. This gives a region
of analyticity which is large enough for most purposes.
For example, it touches the full envelope at points of
interest for the nucleon-nucleon scattering amplitude.
The boundary of an envelope of holomorphy can often be
explored with the help of properly constructed examples
of analytic functions [7].

For problems involving one complex four-vector, and domains of
the form $W\cup R$ considered above, one can obtain the
region of analyticity corresponding to the
envelope of holomorphy of $W\cup N(R)$ with the help of methods from
the theory of distributions and of partial differential
equations. The resulting Jost-Lehmann-Dyson representation
has been discussed widely in the literature [45,46].
It can be viewed as an elegant method to obtain the envelope
of holomorphy for the example we have considered.

As is well known, an elaborate proof of dispersion relations
for amplitudes with fixed values of the momemtum transfer
has been given by Bogoliubov, Medvedev and Polivanov [6].
This proof also makes use of distribution methods and other
tools.

The  actual proof of non-forward dispersion relations starts
with the Fourier representations (4.5). A new variable $\zeta$
is introduced, which corresponds to the squared mass of the
projectile for the actual physical amplitude [47]. For real, and sufficiently
negative values of $\zeta$, the amplitude has cut-plane analyticity
in the energy variable, so that we can write a Hilbert representation.
The problem is then to show that both sides are analytic functions
of $\zeta$, and that the domain of analyticity includes the
physical point $\zeta = \mu^2$. For the left-hand side, we can use the
domain $ E(W\cup N(R))$ discussed before. For the right-hand side, the
required analytic properties can be obtained by a more extensive
use of locality and spectral conditions for the absorptive part
occurring as a weight function in the Hilbert representation.
The tools are again the Edge of the Wedge Theorem, and the envelope
of holomorphy of the domain $D \otimes D$, where $D = W\cup N(R)$ is the
domain described above.

The methods of analytic completion make it possible to prove
dispersion relations for many binary reactions and vertex functions.
For processes like $\pi \pi$- and $\pi N$-scattering, for example,
the proofs are valid for restricted values of the momentum transfer:
$-t = \Delta^2 < \Delta^2_{max} $, with
$$
\Delta^2_{max} = 7\mu^2~~ and~~ \Delta^2_{max} = \frac{8\mu^2}{3}
\frac{2m + \mu}{2m - \mu}
\eqno (4.6)
$$
respectively.\footnote{Tables describing the limitations of proofs
for many amplitudes, which we have prepared for the
1958 Rochester Conference at CERN, are still applicable. See
Ref. 48. The limits are also listed in the appendix of Ref. 49.}
These limitations have no real physical meaning, as can be seen
with the help of models which are unphysical, but satify all the
assumptions we have made [7,31,33].
Nevertheless, it is difficult to incorporate
the information contained in the detailed structure of the intermediate
state spectrum. Of course, the missing features are naturally
contained in a generic, {\it hadronic} perturbation theory, but
in QCD we may not want to rely on that.

There are similar problems for forward scattering amplitudes with
unphysical, continuous contributions. An important example is elastic
nucleon-nucleon scattering, where the envelope of holomorphy
leads to the limitation $\mu > (\sqrt{2} - 1)m $, which is not
satisfied for pion $(\mu)$ and nucleon $(m)$ masses.
The same limitation is obtained for the pion-nucleon vertex
function in the pion channel, and for electromagnetic
form factors of the nucleon. Using
formal perturbation theory simulations, we find that the restriction
is due to singularities describing the composite structure of
the nucleon with respect to physically non-existent particles with
masses such that the simple spectral conditions are satisfied [31,33].
Again, a more exhaustive use of the unitarity condition is required,
but difficult to implement, in particular for intermediate states
with more than two particles. In contrast, dispersion relations
for the pion-nucleon vertex function in a nucleon channel can
be proven using the gap method [4,7]. As we have mentioned, they
are of importance for a complete derivation of the pion-nucleon
relations.

For amplitudes involving strong and electromagnetic interactions,
we may consider dispersion relations involving their hadronic
structure, treating the electromagnetic interaction in lowest,
non-trivial order. Within this framework, we can prove dispersion
relations for pion photoproduction and similar reactions [49].
The limitations in momentum transfer may be found in Refs.49 and
48. There is also no difficulty in deriving a dispersion
representation for the electromagnetic form factor of the pion.

The envelope of holomorphy $E(W\cup N)$ for the amplitudes
$F(K)$ can be used in order to show that the real and
imaginary parts of the corresponding amplitudes are analytic
functions in momentum transfer, or in $cos\theta = 1 - \frac{2t}{K^2}$.
They are regular in the small or large Lehmann-ellipses respectively [50].
Consequently, there are convergent partial-wave expansions.
For the absorptive parts of reactions like $\pi \pi$- or
$\pi N$- scattering, these expansions provide a representation
of the weight function
in the unphysical region, which is always present in dispersion
relations for finite momentum transfer.

Further discussions of pion-nucleon dispersion relations
may be found in the papers [51].

An interesting proposal for the analytic structure of binary
amplitudes has been made by Mandelstam [52]. The double dispersion
relations are essentially based on the assumption that the
singularities of the amplitudes are restricted to those
expected on the basis of physical intermediate states in the three
channels s, t and u, where s + t + u = $\Sigma  m^2$. As is evident
from our previous discussion, these representations have not been
proven in general hadronic field theory, and hence in QCD.
They are known to be compatible with hadronic perturbation theory
in lower orders. As mentioned before, hadronic perturbation
theory may not be a valid approach as far as QCD is concerned.
However, it could provide a hint for the analytic structure
of hadronic amplitudes.

\bigskip

\bigskip

\bigskip

\centerline{ACKNOWLEDGMENTS}
\smallskip

This article owes its existence to the persistent, friendly
persuasion by Gerhard H\"{o}hler. We also would like to thank
Harry Lehmann, Yoichiro Nambu and Wolfhart Zimmermann
for helpful conversations and remarks.
This work has been supported in part by the National Science
Foundation, grant PHY 91-23780.

\newpage
\bigskip
\bigskip

\def\pr{Phys. Rev.~}
\def\nc{Nuovo Cimento~}
\def\np{Nucl. Phys.~}
\def\ptp{Prog. Theor. Phys.~}
\def\pl{Phys. Lett.~}
\def\mpl{Mod. Phys. Lett.~}
\def\mlg{M. L. Goldberger}
\def\wz{W. Zimmermann}
\def\ro{R. Oehme}

\centerline{REFERENCES\footnote{This article touches
many areas of field theory, and it is not feasible to give
a complete list of references. For more extensive lists, we
refer to the review articles covering the various topics
involved.}}

\begin{enumerate}

%1
\item \mlg, H. Miyazawa and \ro, \pr {\bf 99}, 986 (1955);
\mlg, Y. Nambu and \ro, Ann. Phys. (New York) {\bf 2}, 226 (1956);
\mlg, \pr {\bf 99}, 979 (1955).

%2
\item \ro, \pr {\bf 100}, 1503 (1955); {\bf 102}, 1174 (1956).

%3
\item \mlg, Y. Nambu and \ro, reported in: {\it Proceedings of the Sixth
Annual Rochester Conference} (Interscience, New York, 1956) pp.1-7;
G. F. Chew, \mlg, F. E. Low and Y. Nambu, \pr {\bf 106} 1337 (1957).

%4
\item \ro, \nc {\bf 10}, 1316 (1956). \\
\ro, in {\it Quanta}, ed. by P. Freund, C. Goebel and Y. Nambu (Univ. of
Chicago Press, Chicago, 1970) pp. 309-337.

%5
\item K. Symanzik, \pr {\bf 100}, 743 (1957).

%6
\item N. N. Bogoliubov, B. V. Medvedev and M. V. Polivanov,{\it Voprossy Teorii
Dispersionnykh Sootnoshenii} (Fitmatgiz, Moscow, 1958); \\
N. N. Bogoliubov and D. V. Shirkov, {\it Introduction to the Theory of
Quantized Fields} (Interscience, New York, 1959).

%7
\item H. J. Bremermann, \ro~ and J. G. Taylor, \pr {\bf 109}, 2178 (1958).

%8
\item H. Lehmann, Suppl. \nc {\bf 14}, 1 (1959).

%9
\item M. Froissart, in {\it Dispersion Relations and their Connection with
Causality}, ed. by E. P. Wigner (Academic Press, New York, 1964) pp. 1-39.

%10
\item R. Kronig, J. Opt. Soc. Am. {\bf 12}, 547 (1926); \\
H. A. Kramers, {\it Atti del Congresso Internazionale de Fisici, Como}
(Nicolo Zanichelli, Bologna, 1927) p. 545.

%11
\item M. Gell-Mann, \mlg~ and W. E. Thirring, \pr {\bf 95} 1612 (1954).

%12
\item A. Wightman, \pr {\bf 101}, 860 (1956).

%13
\item C. Becchi, A. Rouet and R. Stora, Ann. Phys. (N.Y.) {\bf 98}, 287
(1976); I. V. Tyutin, Lebedev report FIAN No. 39 (1975).

%14
\item T. Kugo and I. Ojima, \ptp Suppl. {\bf 66}, 1 (1979).

%15
\item M. Spiegelglas, \np B{\bf 283}, 205 (1987).

%16
\item K. Nishijima, \np B{\bf 238}, 601 (1984).

%17
\item T. Kugo and S. Uehara, \ptp {\bf 64}, 1395 (1980).

%18
\item N. Nakanishi, \ptp {\bf 62}, 1396 (1979).

%19
\item I. B. Frenkel, H. Garland and G. J. Zuckerman, Proc. Nat. Acad. Sci.
USA {\bf 83}, 8442 (1986).

%20
\item J. J. Figueroa-O'Farrill and T. Kimura, Stony Brook Report
ITB-Sb-88-34.

%21
\item \ro, \pr {\bf D42}, 4209 (1990); \\
\pl {\bf B195}, 60 (1987);~ {\bf B252}, 641 (1990).

%22
\item K. Nishijima, \ptp {\bf 75}, 22 (1986).

%23
\item K. Nishijima and Y. Okada, \ptp {\bf 72}, 254 (1984).

%24
\item \ro, \pl {\bf B232}, 498 (1989).

%25
\item K. Nishijima, \ptp {\bf 77}, 1035 (1987).

%26
\item H. Lehmann, K. Symanzik and \wz, \nc {\bf 1}, 425 (1955);
{}~{\bf 6}, 319 (1957).

%27
\item K. Wilson, \pr {\bf 179}, 1499 (1968).

%28
\item \wz, \nc {\bf 10}, 597 (1958).

%29
\item K. Nishijima, \pr {\bf 111}, 995 (1958).

%30
\item \wz, in {\it Wandering in the Fields}, ed. by K. Kawarabayashi and A.
Ukawa (World Scientific, Singapore, 1987) pp. 61-80.

%31
\item \ro, in {\it Werner Heisenberg und die Physik unserer Zeit},\\ ed. by
Fritz Bopp (Vieweg, Braunschweig, 1961) pp. 240-259; \\
\pr {\bf 121}, 1840 (1961).

%32
\item Y. Nambu, \nc {\bf 9}, 610 (1958).

%33
\item \ro, \pr {\bf 111}, 143 (1958); \nc {\bf 13}, 778 (1959);
\pr {\bf 117}, 1151 (1960).

%34
\item R. Karplus, C. M. Sommerfield and F. H. Wichmann, \pr {\bf 111}, 1187
(1958); L. D. Landau, \np {\bf B13}, 181 (1959);
R. E. Cutkosky, J. Math. Phys. {\bf 1}, 429 (1960).

%35
\item M. A. Shifman, A. I. Vainstein and V. I. Zakharov, \pl {\bf 78B}, 443
(1978).

%36
\item K. Wilson and \wz, Comm. Math. Phys. {\bf 24}, 87 (1972).

%37
\item R. Haag, \pr {\bf 112}, 661 (1958); \\
D. Ruelle, Helv. Phys. Acta {\bf 35}, 146 (1962).

%38
\item \ro, Mod. \pl A{\bf 6}, 3427 (1991).

%39
\item K. Nishijima, \ptp {\bf 80}, 905, 987 (1988).

%40
\item K. Nishijima, \ptp {\bf 74}, 889 (1985).

%41
\item \wz, in {\it 1970 Brandeis Lectures}, eds. S. Deser, M. Grisaru and H.
Pendleton (MIT Press, Cambridge, 1971) pp.395-591;\\ Ann. Phys. {\bf 71}, 510
(1973).

%42
\item H.-J. Borchers, \nc {\bf 15} 784 (1960).

%43
\item H. Jost, Helv. Phys. Acta {\bf 30}, 409 (1957).

%44
\item J. Bros, A. Messiah and R. Stora, Journ. Math. Phys. {\bf 2}, 639
(1961).

%45
\item R. Jost and H. Lehmann, \nc {\bf 5}, 1958 (1957).

%46
\item R. J. Dyson, \pr {\bf 110}, 1460 (1958).

%47
\item N. N. Bogoliubov, B. V. Medvedev and M. K. Polivanov, Lecture Notes,
Institute for Advanced Study, Princeton, 1957.

%48
\item \mlg, Rapporteur, {\it Proceedings of the 1958 Annual International
Conference on High Energy Physics at CERN} (CERN, Geneva, 1958) pp. 207-211.

%49
\item R. Oehme and J. G. Taylor, \pr {\bf 113}, 371 (1959).

%50
\item H. Lehmann, \nc {\bf 10}, 1460 (1958).

%51
\item G. Sommer, Fortschritte der Physik {\bf 18}, 557 (1970); \\
A. Martin, {\it Lecture Notes in Physics} {\bf No.3} \\
(Springer Verlag, Berlin, 1969).

%52
\item S. Mandelstam, \pr {\bf 112}, 1344 (1958).

\end{enumerate}

\end{document}